\documentclass[
  aps,
  pra,
  reprint,
  nofootinbib,
  longbibliography
]{revtex4-2}

\usepackage{graphicx}
\usepackage{amsmath,amssymb,amsfonts}
\usepackage[dvipsnames]{xcolor}

\providecommand{\equalcont}[1]{\thanks{#1}}

\begin{document}

\title{GNSS-free quantum gravity-aided navigation and fine-scale marine surveying with a strapdown quantum gravimeter}

\author{Patrick J. Everitt}
\equalcont{These authors contributed equally to this work.}
\author{Donald H. White}
\equalcont{These authors contributed equally to this work.}
\author{Todd Lyon}
\author{Murat Muradoglu}
\author{Alessandro D'Ortenzio}
\author{Aaron J. Canciani}
\author{Malo Cadoret}
\author{Daniel D. Brown}
\author{David Adams}
\author{Yosri Ben-A\"{i}cha}
\author{Suraj Bijjahalli}
\author{Mojtaba K. Farsani}
\author{Alexander Rischka}
\author{Karandeep S. Gill}
\author{Magdalena Meyer}
\author{Henry W. Orton}
\author{Nicholas P. Robins}
\author{Reuben Symon}
\author{Michael J. Biercuk}
\author{Michael R. Hush}
\author{Stuart S. Szigeti}
\author{Russell P. Anderson}

\affiliation{Q-CTRL, Sydney, NSW Australia}

\begin{abstract}

Global navigation satellite systems (GNSS) are frequently disrupted or unavailable in maritime settings, and unaided inertial navigation systems (INS) drift without external correction, motivating the development of new approaches to improve navigational accuracy. 
Gravity-anomaly map matching powered by quantum sensing offers a passive, infrastructure-free aid, but field trials demonstrating truly GNSS-free quantum gravimetric navigation have not previously been reported. 
Here we perform gravity map matching and fine-resolution gravity survey using a mobile quantum gravimeter on board a 29 m surface vessel. 
The quantum sensor is hybridized with a classical accelerometer for atom-referenced bias stabilization, and incorporates independent mechanization of a navigation-grade inertial measurement unit.
It is installed in an uncontrolled passenger cabin with no special environmental stabilization, and without any form of reference calibration employed or required. 
Operated in both gimbaled and strapdown configurations over identical vessel traversals, the sensor was used to correct the inertial solution over an 83 km maritime trajectory by referencing locally measured gravity to a satellite-derived anomaly map. 
Gravity-aiding constrains the drift of the GNSS-free inertial solution and delivers bounded positioning, achieving a distance root-mean-square (DRMS) position error at the nautical-mile level.
In this navigational demonstration GNSS is excluded throughout the entire measurement chain, INS mechanization, map-matching protocol, and tilt-correction protocol. 
In a separate GNSS-referenced mode, we use the same system to perform a fine-spatial-scale maritime gravity survey over multiple repeated coastal routes in conditions up to Sea State 4. 
We achieve mGal-level agreement with various gravimetric maps, and sub-mGal repeatability and in-run stability across repeated traversals, with gimbaled and strapdown operation performing comparably over the sampled sea states.
Resolved gravity anomalies reach an along-track scale of $\sim300$ m, some $50\times$ finer than the satellite map's own half-power wavelength. 
An uninterrupted 56 h stationary measurement shows that atom referencing stabilizes the hybrid gravity estimate to a long-term drift approximately $70\times$ lower than the classical channel alone.
These results provide the first same-instrument comparison of gimbaled and strapdown mobile quantum gravimetry and, to our knowledge, the first fully GNSS-independent gravity-map-matching navigation demonstration using a quantum gravimeter, establishing a path toward compact, autonomous-platform-ready quantum sensing for navigation and survey in GNSS-denied maritime environments.

\end{abstract}

\maketitle

\section*{Introduction} %Reviewed by Mike, keep edits small. 

Global navigation satellite systems (GNSS) provide passive, globally accessible position fixes, but their radio-frequency signals cannot be used underwater and remain vulnerable to disruption and deception~\cite{HofmannWellenhof2008,Montenbruck2017,RoyalAcademy2011,Paull2014}. These limitations motivate complementary navigation methods for missions in which GNSS is unavailable or cannot be relied upon continuously. Inertial navigation systems (INS) are ubiquitous, passive, self-contained, and independent systems in deployment, but are insufficient by themselves; inertial-sensor errors rapidly accumulate unless the navigation solution is periodically corrected by an external aid~\cite{TittertonWeston2004}. In a maritime setting, acoustic positioning, Doppler-velocity measurements and sonar-based terrain navigation can constrain this positioning-error growth, but their performance depends on external transponders, active acoustic interrogation, bottom lock or sufficiently informative mapped terrain~\cite{Paull2014}. These approaches remain valuable, but no single one reproduces the availability and operational simplicity of GNSS across all maritime settings. As a result there is significant interest in developing complementary aids to support high-accuracy positioning for circumstances when external infrastructure, acoustic observability, or terrain information is limited. 

One such aid is geophysical navigation, which passively provides position information by matching locally-measured natural fields to preloaded maps~\cite{Rice2004Geophysical}. On board sensors perform observations of the local geophysical field independent of any external signal and without requiring any signal emission, making the general strategy well suited to locations where alternative approaches fail or where stealth is prioritized.  Q-CTRL recently demonstrated this architecture with quantum-assured magnetic-anomaly navigation (MagNav), demonstrating bounded positioning in airborne trials irrespective of navigation-path duration, and outperforming a velocity-aided INS by orders of magnitude~\cite{Muradoglu2025}. Gravity supplies a complementary geophysical observable that is particularly well suited to maritime operation. First, gravitational anomalies tend to vary on a spatial length commensurate with the slower traversal speed of maritime vessels.  Next, gravitational map coverage in marine settings is often superior to available magnetic maps.   Satellite altimetry provides globally gridded marine gravity maps~\cite{Sandwell2021,Yu2024SWOT,SIO_SWOT04_2026}, whereas public magnetic compilations such as EMAG2v3 combine heterogeneous satellite, airborne and shipborne measurements and retain substantial oceanic data gaps~\cite{Meyer2017EMAG2}. 

Gravity-aided navigation (GravNav) has a long but sparsely-documented experimental history~\cite{Moryl1998UGM,Ryan1999Memphis,Rice2000Marine,Rice2004Geophysical, USNavyGAINS2009}, often due to classification of the underlying data. Open surface-vessel trials provide quantitative demonstrations of gravity-aided navigation with classical gravimeters, but the underlying evidence suffers from nonuniform and sometimes obscured levels of GNSS independence~\cite{Wang2016,Wang2024Triangulation,Zhou2026Calibration}.  True mission relevance in GNSS-denied environments of gravity-aided navigation can only be shown once GNSS is unambiguously excluded from the processing chain, including the calculation of E\"otv\"os corrections and attitude.  Further work is thus required to openly demonstrate that GNSS-free GravNav can deliver mission-relevant performance, and one key requirement to overcome historical limits is to improve the overall capabilities of the gravity-measuring device and the associated process of calculating a gravity-aiding correction to the navigation solution.

\begin{figure}[tb]
    \centering
    \includegraphics[width=\columnwidth]{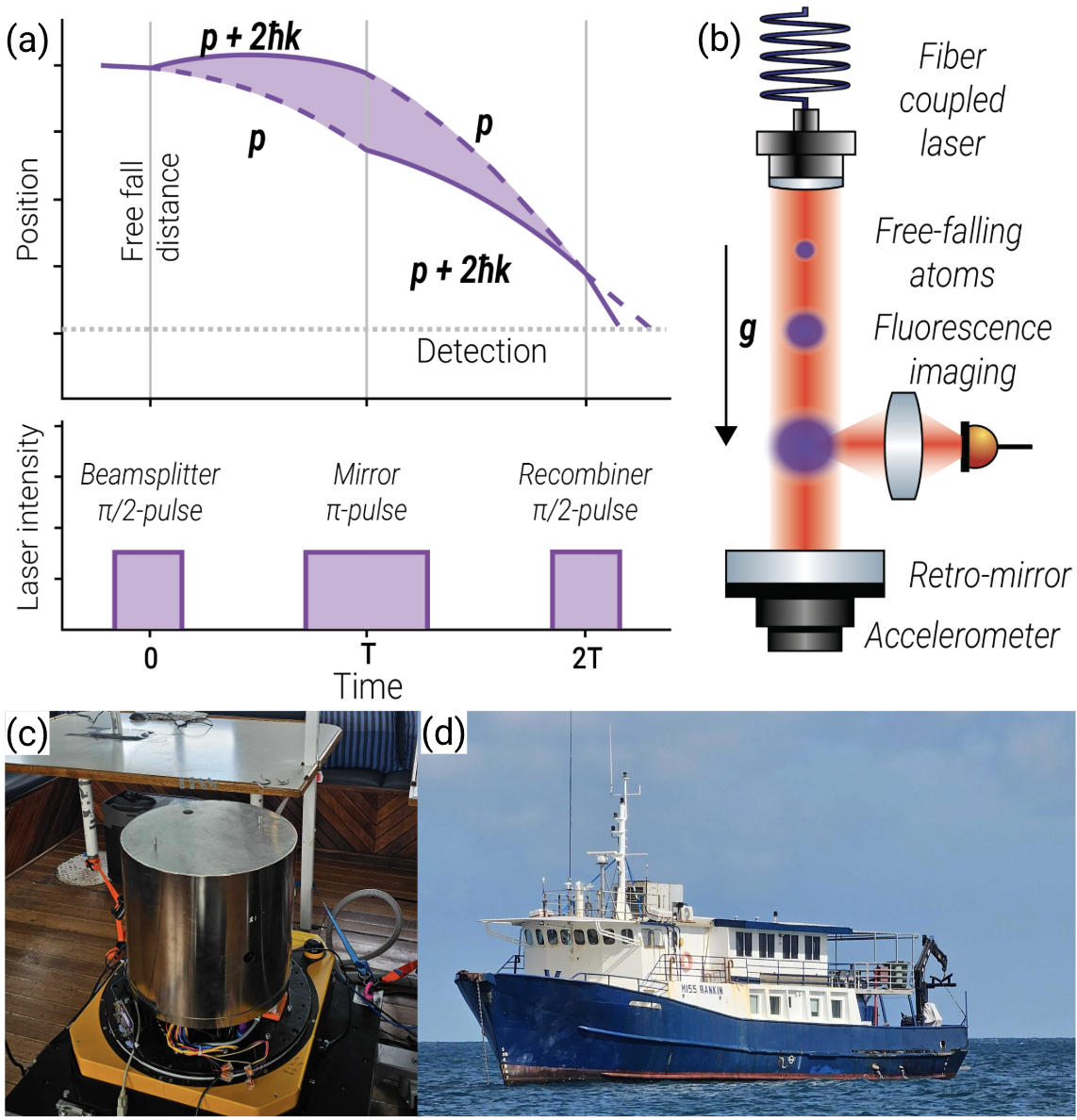}
    \caption{Marine trial with a quantum gravimeter.
    (a), Space--time diagram of the free-falling Mach--Zehnder atom-interferometer sequence used during the trial. Three square laser pulses split, redirect and recombine the $^{87}$Rb matter waves by imparting the atoms with differing momenta $p$ augmented by the quantum of momentum carried by the laser light $\hbar k$; the lower panel shows the corresponding laser-intensity sequence used to perform an acceleration-sensitive interferometer over a total time of $2T$ s.   The shaded area between the effective interferometer arms represents the enclosed interferometer phase imparted due to differential impact of gravity on atoms with different momenta along the gravitational axis.  Detection at the end of the sequence provides a direct readout of the atomic populations and the phase of the interference pattern, providing a local probe of gravity. 
    (b), Sensor-head schematic. A fiber-coupled interferometry beam interrogates the freely falling atoms, and the output-state populations are measured by fluorescence imaging (FI). The beam is retro-reflected from a mirror carrying the classical accelerometer used for hybridization.
    (c), Quantum gravimeter installed in the vessel's top-deck cabin, described previously in~\cite{Everitt2025}. The sensor head was mounted on a lockable gimbal (orange platform) to enable alternating measurements between gimbaled and strapdown operation; the laser and control-electronics rack is visible at upper right and connected via umbilical to the sensor head.
    (d), \textit{MV Miss Rankin}, the 29 m-long, 7.3 m-beam vessel used for the maritime field trial.}
    \label{fig:marine-trial}
\end{figure}

\begin{figure*}[tb]
    \centering
    \includegraphics[width=\textwidth]{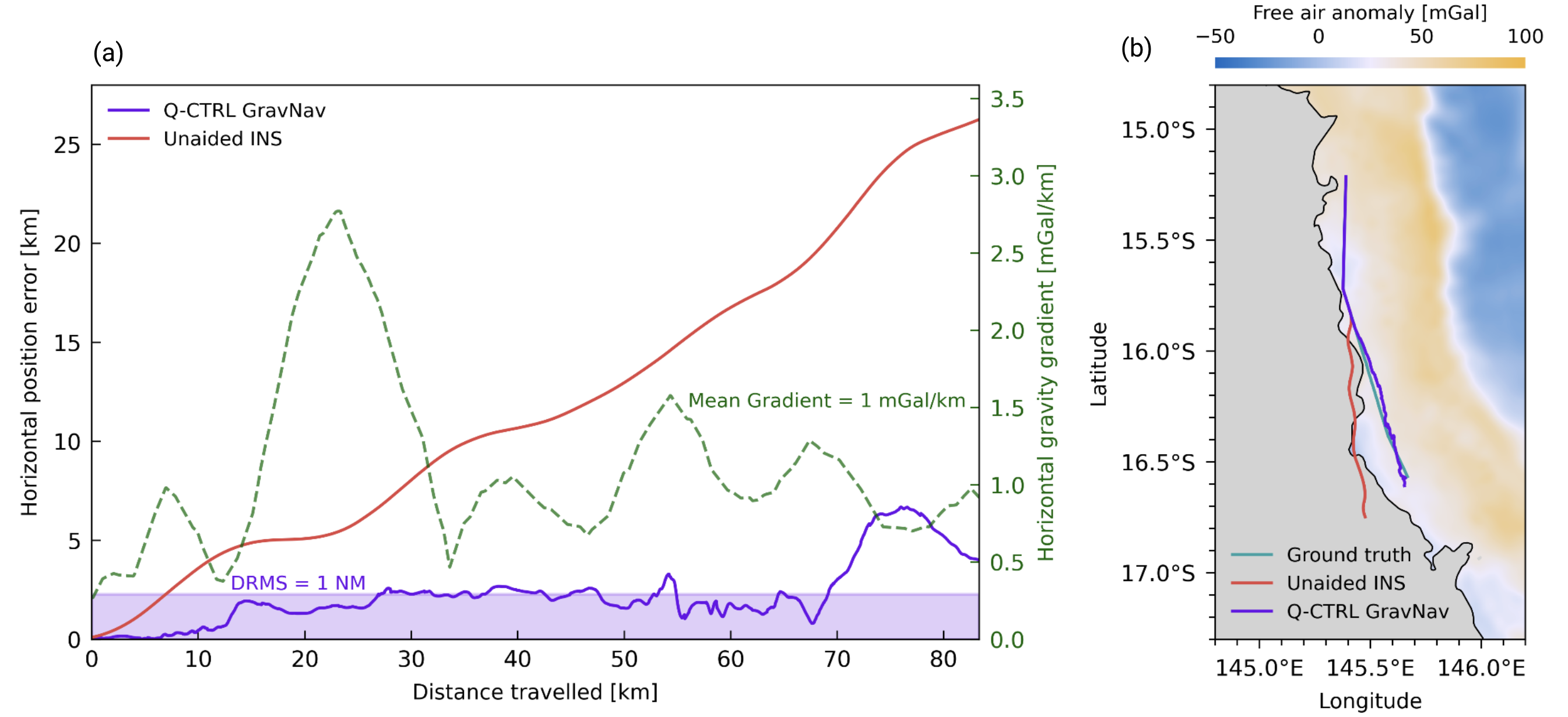}
    \caption{Demonstration of gravity-aided navigation using a quantum gravimeter.
    (a), Pointwise horizontal position error versus distance traveled for Q-CTRL GravNav (purple) and the unaided inertial navigation system (INS; red). Both solutions exclude GNSS observations during the evaluated segment; GNSS is used only as the post-processed reference against which error is calculated. The dashed green trace, on the right axis, is the magnitude of the horizontal gradient of the grav\_SWOT\_04 map evaluated along the withheld ground-truth trajectory and is diagnostic only. The purple band marks horizontal errors below one nautical mile (1.852 km); it is a distance reference rather than a reported distance root-mean-square (DRMS) statistic. 
    (b), Ground-truth trajectory (teal), unaided INS trajectory (red), and Q-CTRL GravNav trajectory (purple), overlaid on the grav\_SWOT\_04 free-air gravity anomaly map~\cite{SIO_SWOT04_2026,Yu2024SWOT,Sandwell2021}.  Grey depicts land in this map, such that the free-air gravity anomaly is only represented for the marine environment.}
    \label{fig:gravnav}
\end{figure*}

Quantum gravimeters offer two potential advantages for this task: (i) an intrinsically referenced, low-drift (low-environment-sensitivity) measurement that can reduce reliance on operational bias calibration; and (ii) a pathway to lower size, weight and power (SWaP) for deployment on autonomous platforms. In an atom interferometer, the acceleration scale factor is set by the effective optical wavevector and pulse timing rather than by the calibration of a mechanical proof mass~\cite{Bidel2013Field,Bidel2020Airborne,Bidel2023Airborne}.  Repeated preparation of freely falling atoms provides a low-frequency reference that can stabilize the bias of a continuously sampled, higher-bandwidth classical accelerometer~\cite{Menoret2018AQG,Wu2019MobileAtom,Lautier2014Hybrid,Cheiney2018Hybrid,Bidel2018}.  These devices, when compared against gravity reference data, consistently show that they can be approximated as ``calibration-free'' absolute measurements of gravity (i.e. the process of measurement transduction does not require any form of periodic calibration against an external reference when functioning correctly). In contrast, classical maritime gravimeters can compensate drift effectively---for example, an estimated 6 mGal linear drift over two months can be corrected to approximately 1--2 mGal---but doing so typically requires absolute gravity ties, empirical drift models, strict environmental controls, or repeated-line and crossover adjustment~\cite{Wang2017MapCharacteristics,Cai2017Strapdown,Lu2022Marine,SOESTBGM3}.   For SWaP, U.S. Department of War programmes target complete quantum-gravimeter packages of 10--20 L under DARPA RoQS~\cite{DARPARoQS2026} and DIU FARSEER~\cite{DIUFarseer2026}, substantially below the 33--134 L packaged volumes reported for high-performing classical marine strapdown systems~\cite{Johann2023Thesis,Wang2018SGAWZ02,Wang2018Strapdown}.

Despite their potential advantages and significant simulation-based studies~\cite{Phillips2020AugmentedINS,Phillips2022PositionFixing,Lellouch2025QuantumGradiometry}, field demonstrations of gravity-aided navigation using a quantum sensor are notably absent from the open literature.  Work to date has focused on GNSS-aided mobile quantum gravimetric survey and passive detection of spatial mass anomalies. These experiments demonstrated accurate quantum-gravity recovery under way, but the reported maritime implementations used mechanical stabilization for the sensitive axis and GNSS for dynamic corrections or georeferencing. The same group's earlier field-portable and airborne quantum gravimetry demonstrations relied on GNSS in the same way~\cite{Bidel2013Field,Bidel2020Airborne,Bidel2023Airborne}.  

In this manuscript we report a field-deployed quantum gravimeter used for: (i) true GNSS-free gravity-aided navigation on a maritime surface vessel (Fig.~\ref{fig:marine-trial}) and (ii) strapdown maritime gravity survey under way. First, quantum-stabilized acceleration and gravity-map matching are used to correct raw measurements from a manufacturer-classified ``navigation-grade'' fiber-optic-gyroscope (FOG) inertial measurement unit (IMU) (Advanced Navigation Boreas D90), while GNSS is withheld from the complete measurement, inertial-mechanization, gravity-recovery, map-matching, and state-correction chain~\cite{AdvancedNavigationBoreasD90,AdvancedNavigationINSGrades}. We demonstrate gravity-map-aided position correction from a quantum gravimeter over a 6 h, 45 nmi (83 km) marine trajectory by independently mechanizing the raw angular-rate and specific-force outputs of the IMU. 14 nmi (26 km) to 2.2 nmi (4.1 km), a factor of $\sim 6.3 \times$  More importantly, the GravNav demonstration shows bounded positioning of $1$ nautical mile over the majority of the traversal. 

In order to gain deeper insights into the achieved performance of the gravity-aided navigation, we separately operate the same instrument in a GNSS-referenced surveying mode to test its recovery of repeatable gravity and spatial anomalies from the same moving maritime platform. These measurements resolve repeatable 15--40 mGal features at an along-track scale of approximately 0.17 nmi (0.3 km). Independent GNSS-referenced measurements enable a direct within-instrument comparison of gimbaled and strapdown quantum-gravimeter operation, a comparison not yet presented in previous mobile-quantum-gravimetry demonstrations.  We achieve an in-run stability of 0.67 mGal and a run-to-run repeatability of 0.16 mGal in strapdown operation, demonstrating the efficacy of a range of software-ruggedization techniques employed in sensor operation and signal processing.  Finally, we demonstrate mobile quantum gravimetry in a road trial, achieving continuous quantum gravimetry at 100 km h$^{-1}$ with repeatability $\sim6.3$ mGal. Together, these results enable independent evaluation of quantum gravity measurement and GNSS-free GravNav within a single field-deployed sensor, without relying on cross-referencing of measurements, a first in the community.  

\begin{figure*}[tb]
    \centering
    \includegraphics[width=\textwidth]{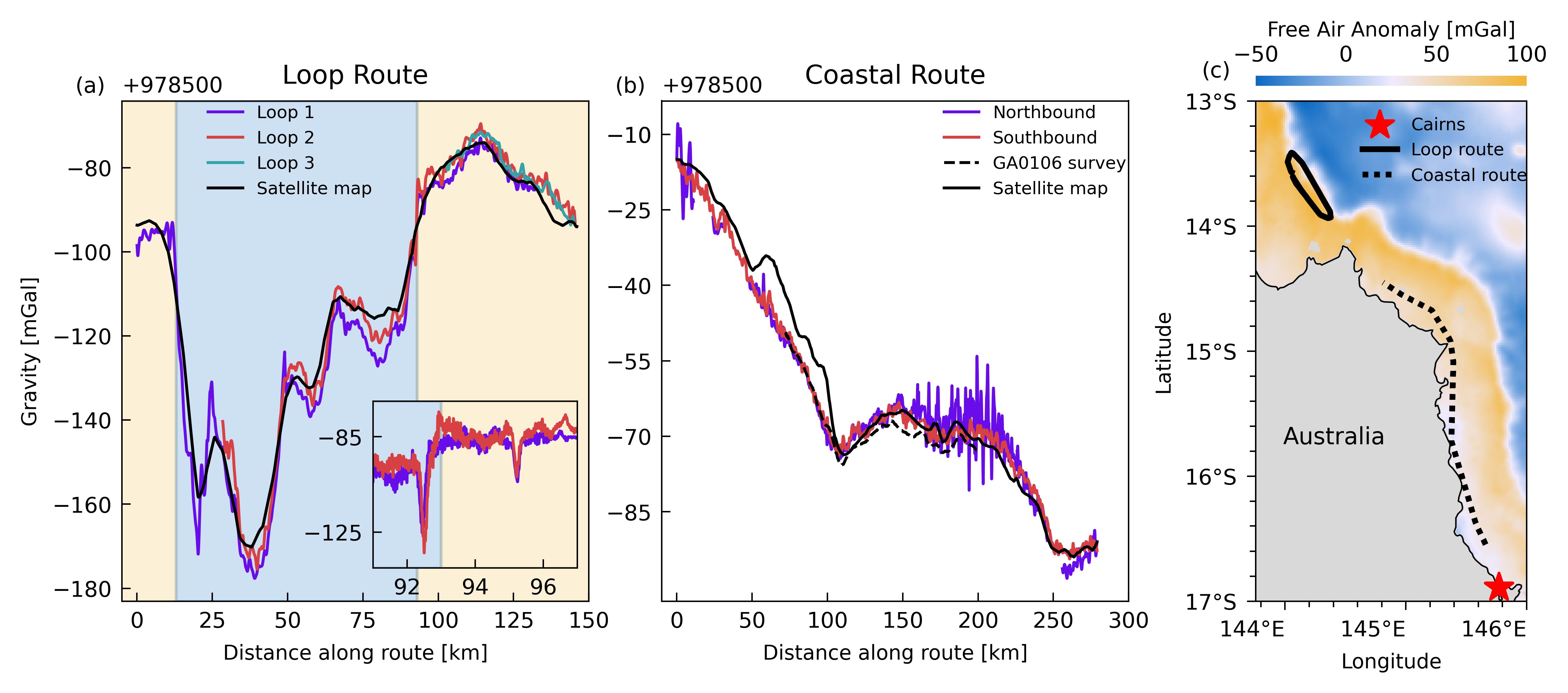}
    \caption{Comparison of quantum-gravity measurements with satellite and ship-survey gravity.
    (a), Gravity measured during three traversals of the 145 km loop route in (c). A Bessel low-pass filter with a 300 s time constant is applied, and gravity predicted by grav\_SWOT\_04 is shown in black~\cite{SIO_SWOT04_2026,Yu2024SWOT,Sandwell2021}. The inset expands the 92--96 km region using a 60 s filter. Kilometer zero is $(13.5725^{\circ}\mathrm{S},144.0567^{\circ}\mathrm{E})$, and distance is measured clockwise from this point. Blue shading from 13 to 93 km denotes open water outside the Great Barrier Reef; yellow shading denotes water within the reef. Loops 1 and 2 are clockwise and Loop 3 anticlockwise. Mean vessel speed is 8.4 knots.
    (b), Gravity measured on northbound and southbound traversals of the 280 km coastal route in (c). Kilometer zero is the southern endpoint at $(16.5482^{\circ}\mathrm{S},145.6611^{\circ}\mathrm{E})$. The northbound and southbound traces were each filtered using a Bessel low-pass filter with a 300 s time constant. grav\_SWOT\_04 and the nearby GA-0106 seaborne survey line from the Geoscience Australia Mardat archive~\cite{Petkovic2001} are overlaid. GNSS-derived position and velocity were used for this gravimetry analysis.
    (c), The 145 km loop and 280 km coastal trajectories overlaid on the grav\_SWOT\_04 free-air gravity-anomaly map.}
    \label{fig:gravity-surveys}
\end{figure*}

\begin{figure*}[t]
    \centering
    \includegraphics[width=\textwidth]{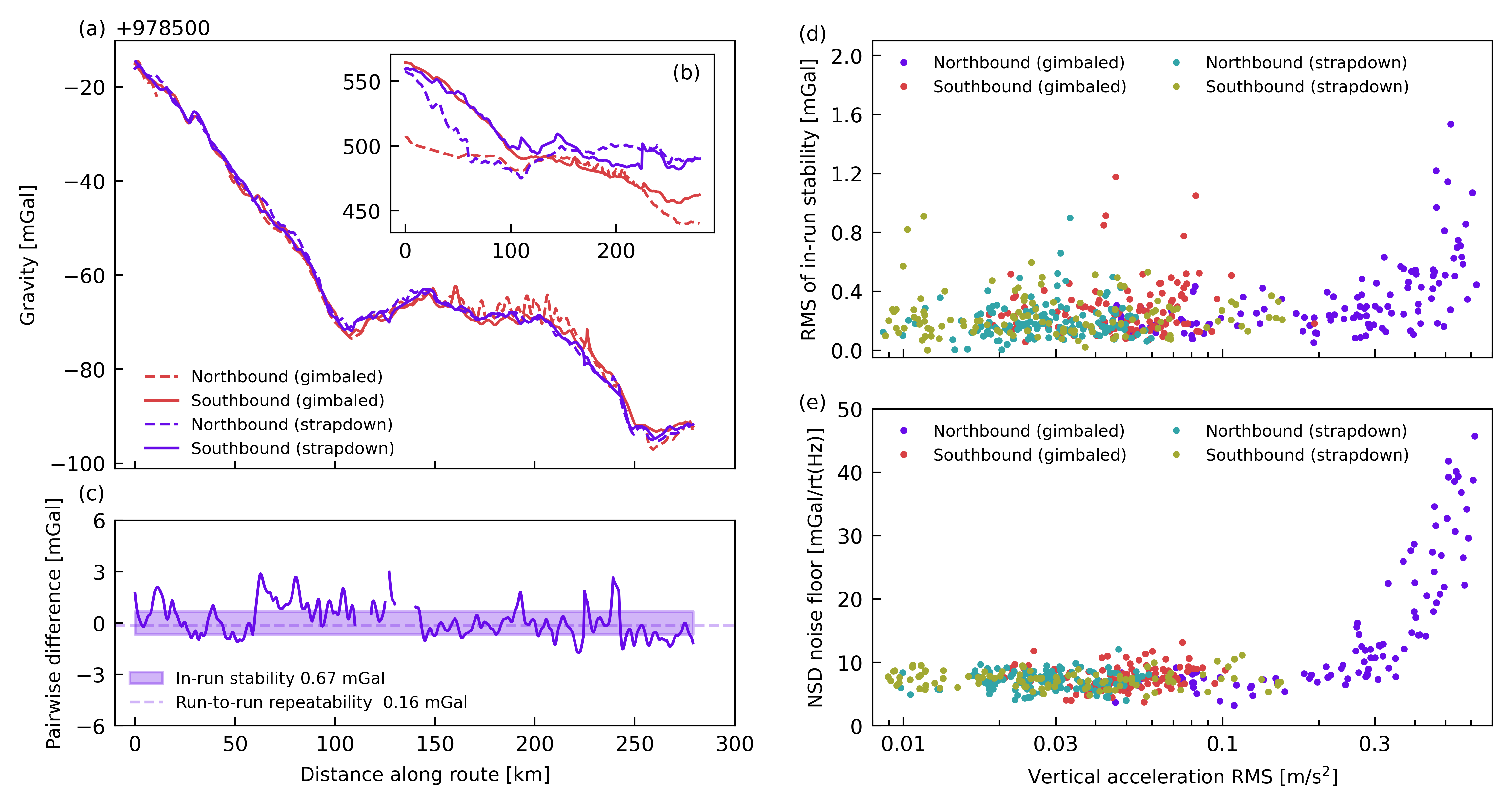}
    \caption{Comparison of gravimeter performance in gimbaled and strapdown configurations.
    (a), Gravity measured along the 280 km coastal route during northbound and southbound traversals with the quantum gravimeter operated in gimbaled and strapdown configurations.
    (b), Gravity estimates along the same 280 km coastal route from the same gravimetric pipeline without hybrid quantum-classical stabilization applied, relying exclusively on classical sensors (see \emph{Methods}).
    (c), Effective pairwise residual between the repeated gravity profiles as a function of distance. The shaded band denotes the route-wide residual RMS of 0.67 mGal and the dashed line the run-to-run repeatability of 0.16 mGal.
    (d), RMS of the mean pairwise sensor in-run stability evaluated over 1000 s windows versus RMS out-of-band sea-state-induced vertical acceleration. Vertical acceleration is used as a measured proxy for vessel-motion severity. Colors identify the northbound and southbound gimbaled and strapdown runs.
    (e), Low-frequency amplitude noise spectral-density (single-sided) floor below 0.01 Hz, evaluated over the same 1000 s windows. The two configurations have overlapping residual and spectral-noise distributions over their shared low-to-moderate-motion range; the increased residual and noise at the largest accelerations are confined to the northbound gimbaled run, the only run that samples those conditions.}
    \label{fig:gimbal-strapdown}
\end{figure*}

\section*{Results}

We operate the quantum sensor aboard the 29 m research vessel \textit{MV Miss Rankin} (Fig.~\ref{fig:marine-trial}d) from a top-deck cabin (Fig.~\ref{fig:marine-trial}c). The same instrument is used for the navigation, gravity-survey and configuration-comparison measurements, with a lockable gimbal permitting both gimbaled and rigid strapdown operation (see \emph{Methods}, ``Field trials'').  

The installation location is enclosed to protect against weather and sea spray, but is otherwise open to the environment. Neither the sensor head nor its flight case incorporates active or passive temperature control or a dedicated environmental enclosure; active temperature regulation is confined to small subcomponents within the laser system. All measurements reported below are retrieved from the instrument in this location across a range of conditions up to Sea State 4.  For full details on the trial traversals, sea-state conditions, and vessel accelerations experienced please see \emph{Methods}.

\subsection*{Quantum navigation with gravity map matching} 

We evaluate the complete gravity-aided navigation chain on a 45 nmi (83 km) segment of a southbound coastal route (Fig.~\ref{fig:gravnav}, see \emph{Methods}, ``Gravity-aided navigation''). The measured gravity sequence is matched to the preloaded grav\_SWOT\_04 free-air anomaly map~\cite{SIO_SWOT04_2026,Yu2024SWOT,Sandwell2021}, and the resulting information is used to correct the inertial state. All GNSS observables are withheld from the inertial mechanization, gravity recovery, map matching and state-correction chain, and the recorded GNSS solution is used only after processing to construct the ground-truth trajectory in Fig.~\ref{fig:gravnav}b,c. 

We report two navigation solutions, both initialized from the same state: the unaided (drifting) INS and the GravNav-aided positioning. Position error is calculated as the separation between each estimated position and the GNSS ground-truth. As anticipated, the unaided INS error increases due to sensor drift to approximately 14 nmi (26 km) at the end of the route. In contrast, the Q-CTRL GravNav error ended at 2.2 nmi (4.0 km), corresponding to an endpoint error reduction of approximately $\sim 6.3 \times$. More importantly, over the first 38 nmi (70km) of the traversal, the GravNav positioning error is bounded, exhibiting $1$ nmi error while the INS accumulates unbounded drift.  This difference is material as Fig.~\ref{fig:gravnav}c shows that the GravNav solution maintains a relatively accurate traversal path whereas the drifting INS reports the vessel running aground.

The green dashed line in  Fig.~\ref{fig:gravnav}b shows the horizontal gravity gradient over the traversal.  As observed from these data, the aiding map itself is spatially band-limited. Along-track spectral density analysis of grav\_SWOT\_04 shows that it reaches 50\% of cumulative anomaly power at a wavelength of approximately 16 km, and 95\% of its power lies above $\lambda_{95\%}\approx6$ nautical miles (11.1 km) . The 4.1 km terminal position error achieved without GNSS is therefore already smaller than this $\lambda_{95\%}$ band-limit, by a factor of approximately $2.7\times$, despite the map carrying little power below this wavelength. As shown independently below (Fig.~\ref{fig:gravity-surveys}), the same gravimeter recovers real gravity structure at least $50\times$ finer than this map's own half-power wavelength.

\subsection*{Repeated routes recover fine gravity anomalies} 

We next focus on the application of our system as a mobile gravimeter in an effort to gain insights into the process by which a gravity-based aid is extracted and the potential limits of this process across a broad range of conditions. Extracting gravity from measured accelerations in a mobile system relies on external aiding to obtain velocity-dependent E\"otv\"os corrections, as well as the sensor attitude in order to perform coordinate transforms to the Earth frame. In this section we use GNSS aiding for these tasks, with the goal of understanding the repeatability, the sensitivity and the accuracy of the system under varying conditions at sea. We process this data independently and supply no information to the GNSS-free GravNav result. The two routes in the Coral Sea shown in Fig.~\ref{fig:gravity-surveys}(c) are traversed multiple times (see \emph{Methods}), enabling direct comparisons of sensor performance over repeated tracks.

The loop route has rich gravitational variation owing to its position at the edge of the continental shelf. Figure~\ref{fig:gravity-surveys}(a) shows that repeated traversals of the route exhibit a high degree of repeatability, including in Sea State 4 conditions outside of the reef. The measured quantum-gravity survey results generally follow the grav\_SWOT\_04 satellite map at the map’s resolution. Notably, we observe a number of features at a fine scale well beyond the resolution of the map. In particular, the section shown in the inset to panel (a) between 91-97 km shows two features resolved in both Loop 1 and Loop 2 which are approximately 300 m wide. The features are detected in the two traversals with repeatable positions and anomaly heights, while being absent from the satellite map.

%By contrast, removing the quantum sensor from the gravity recovery process while using the same classical accelerometer/ FOG infrastructure delivers results with repeatability errors more than an order of magnitude larger than when hybridizing the atomic gravimeter. 

Figure~\ref{fig:gravity-surveys}(b) compares the gimbaled traversals of the coastal route with the satellite map. Variation with the satellite map, particularly in the section 50-100 km, is attributable to the proximity to the coastline~\cite{Zampa2022}. This is supported by comparison with the GA-0106 maritime survey~\cite{Petkovic2001,NOAANCEI2026} which followed a similar route and agrees well with the 50-100km measurements. This local preference for the ship-survey trace is consistent with known degradation of altimeter-derived gravity near coastlines: independent comparisons have identified land-contamination artefacts extending 17 km offshore~\cite{Zampa2022}, and a previous quantum-gravimeter survey found a 7.2 mGal offset in a coastal satellite model~\cite{Bidel2018}.
The northbound gimbaled trace is noticeably noisier in the section between 150-225 km. We attribute this to the sea state in this section of the trial, including potential issues with the gimbal hitting rails for maximum deflection, and we explore this in further detail in the following section.

\subsection*{Strapdown operation preserves repeatability over the sampled sea states}
We now compare the performance of the sensor in gimbaled and strapdown operation, leveraging the ability to rigidly lock the gimbal mount and rely instead on software corrections to account for platform motion. The 280 km coastal route is inside the reef along the full length, and is traversed twice with the sensor gimbaled, and twice with the sensor in strapdown mode, for a total of four traversals (see \emph{Methods}). 

Figure~\ref{fig:gimbal-strapdown}(a) shows that the strapdown measurements notably follow the gimbaled measurements without systematic bias or appreciable added high-spatial-frequency variability. The importance of the quantum sensor in the processing chain is highlighted in Fig.~\ref{fig:gimbal-strapdown}(b), which shows the four traces analyzed only with the onboard classical sensors. The variation between the classical-only traces is significant, together with an overall offset of 570 mGal, linked to classical sensor bias and drift.

Independent benchmarks of global altimetry-derived gravity products have uncertainties of several mGal and show degradation near coastlines~\cite{Li2022AltimetryReview,Zampa2022}. For this reason, pairwise differences at common locations between the four traces are used to obtain repeatability metrics instead of map residuals. Figure~\ref{fig:gimbal-strapdown}(c) shows the pairwise difference between the two strapdown traces, with an in-run stability of 0.67 mGal at 1200 seconds, and an overall mean difference of 0.16 mGal.   Using a shorter integration time of 300s the in-run stability remains at a comparable scale, reaching just 0.93 mGal.  This sub-mGal repeatability -- even when comparing gimbaled and strapdown operation -- is the principal result of this gravimetry work.

Figure~\ref{fig:gimbal-strapdown}(d) looks at the dependence between the RMS of the ensemble of pairwise differences for the four different traces as a function of heave. Heave (vertical acceleration) is used as a simple proxy for ship motion, and larger heave is correlated with larger rotation rates, tilts, surge and sway. Below 0.3 m s$^{-2}$ RMS heave, the in-run stability is independent of motion at an average 0.23 mGal. The northbound gimbaled trace experiences the most unsettled ocean conditions, as it is obtained in the aftermath of a tropical low. The northbound gimbaled trace above 0.3 m s$^{-2}$ RMS heave exhibits worse RMS in-run stability, as seen especially between 150-225 km in Fig.~\ref{fig:gimbal-strapdown}(a).The pairwise differences also allow us to obtain an estimation of the in-motion noise floor from the amplitude noise spectral density below 0.01 Hz. Below 0.3 m s$^{-2}$ RMS heave, the noise spectral density gives an average spectral noise floor of 7.46 mGal/$\sqrt{\text{Hz}}$. In combination these results indicate that over a 30dB amplitude range in heave we maintain a flat effective noise floor for the quantum gravimeter.

\begin{figure}[tb]
    \centering
    \includegraphics[width=\columnwidth]{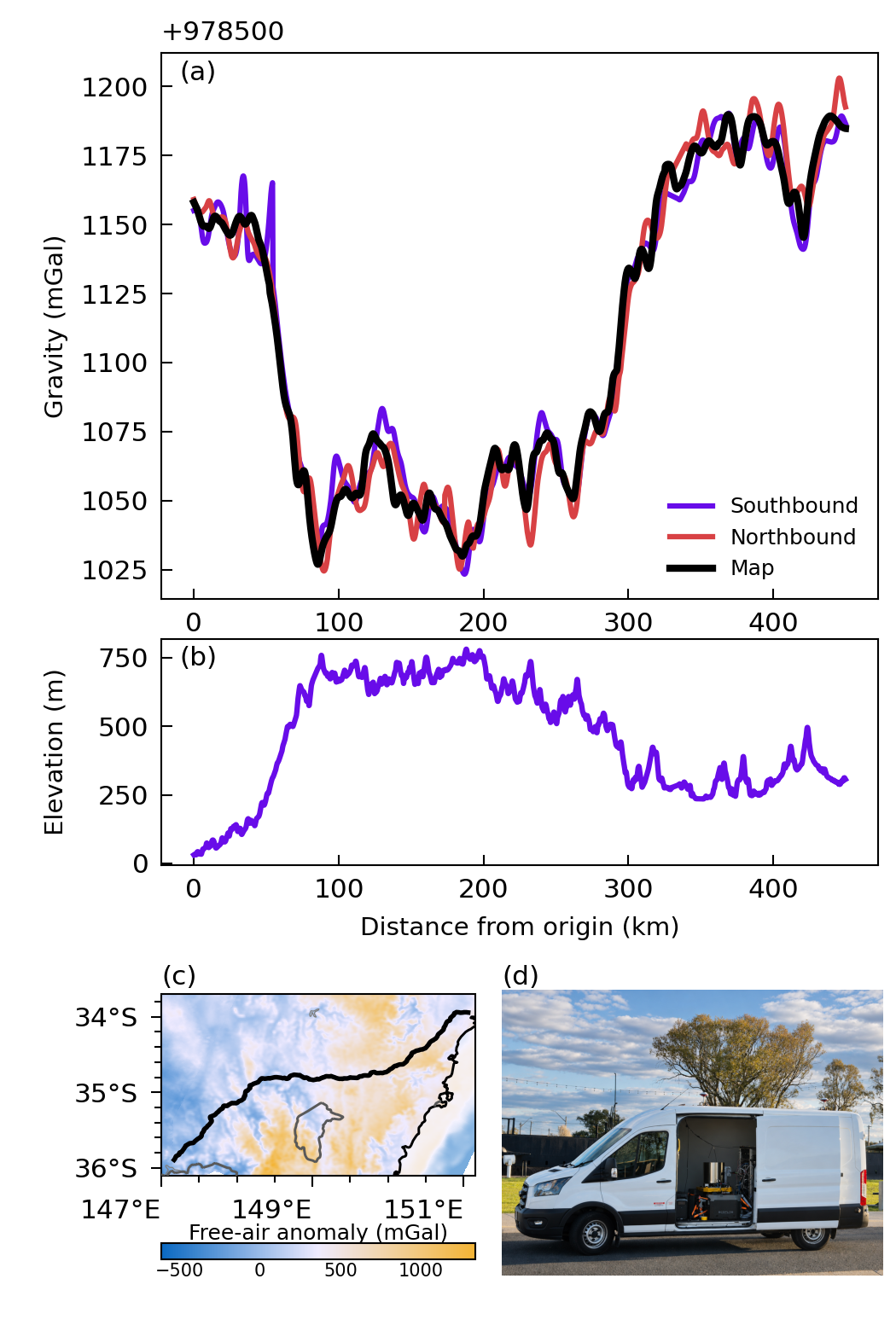}
    \caption{Road gravimetry survey along the Hume Highway in New South Wales.
    The quantum gravimeter was mounted on a gimbal inside a van and driven along the highway at an average speed of 100 km h$^{-1}$. 
    (a) Northbound (red) and southbound (purple) gravity profiles are compared with the gravity trace extracted from an airborne-survey map (black). Kilometer zero is the north-eastern endpoint in south-west Sydney; distance increases towards Mullengandra in southern New South Wales. 
     (b) WGS84 elevation along the journey acquired by GNSS. 
    (c) The route from Sydney to Mullengandra is plotted on the free-air anomaly map from the airborne gravimetry survey~\cite{SanderSurvey2024}. 
    (d) The gravimeter mounted on the gimbal in the van is pictured at a rest stop along the route.}
    \label{fig:road-gravity}
\end{figure}

\subsection*{Highway operation recovers regional gravity structure at 100 km h$^{-1}$}. 

On a separate trial, we strap the sensor in a van and drive on a 450 km route along the Hume Highway in New South Wales with an average speed of 100 km h$^{-1}$, performing gravity survey in motion at highway speeds (Fig.~\ref{fig:road-gravity}(a)). The sensor operates in a gimbaled mode and the route driven is shown in Fig.~\ref{fig:road-gravity}(c), again with no special environmental controls or conditioning in the van's cargo hold (shown in Fig.~\ref{fig:road-gravity}(d)). This survey provides a very different operational environment to a maritime trial and additionally includes significant elevation changes of 750 m along the route (shown in Fig.~\ref{fig:road-gravity}(b)). This necessitates an additional processing step of subtracting the second time derivative of the elevation ($\ddot{h}$) from the vertical acceleration, in order to distinguish between gravitational variation and acceleration due to elevation change. The heave on this trial varied between 0.2-0.78 m s$^{-2}$ with an RMS heave of 0.37 m s$^{-2}$.

Figure~\ref{fig:road-gravity}(a) shows the measured gravity signal compared to an airborne gravimetry survey~\cite{SanderSurvey2024}, indicating good agreement both qualitatively and quantitatively.  The survey reproduces the high-level variation correlated with altitude change over the route and also reveals fine km-scale features present in the map estimate of gravity.  The northbound and southbound traces exhibit an RMS difference from the map of 7.7 mGal and 8.2 mGal respectively and mean offsets of 0.12 and -0.4 mGal respectively. The pairwise difference between the northbound and southbound traces exhibits an in-run stability of 6.3 mGal and a mean difference of 0.52 mGal. 

\section*{Discussion}

The central result of this work is that a gravity sequence measured by a quantum sensor can be used to correct an inertial navigation solution without any GNSS observations, and deliver bounded positioning in a dynamic maritime environment. In order to support this, we present both a direct truly GNSS-free GravNav demo and a series of mobile gravity-survey demonstrations which provide detailed insight into how the gravity aiding delivers improved navigational performance.  

Over an approximately 6 h, 45 N Mi (83 km) maritime evaluation, gravity aiding reduced the horizontal error of the independently mechanized INS to lie within a bound of approximately 1 nmi.  In our demonstrations the measured IMU data, quantum-gravity recovery, map matching and inertial-state correction form an explicitly GNSS-independent chain; GNSS is withheld until the resulting trajectory is evaluated and used only as a ``ground truth'' benchmark.  

These results can be compared against a limited number of published classical-benchmark demonstrations. The Universal Gravity Module combined a classical gravimeter, gravity gradiometers, stored maps and an INS, and processed four observation classes: gravity-anomaly map matching, gravity-gradient map matching, gravimeter--gradiometer comparison to observe velocity error through the E\"otv\"os effect, and gravity-gradient-derived vertical deflection~\cite{Moryl1998UGM,Rice2000Marine,Rice2004Geophysical}. Trials aboard USS \textit{Memphis} in 1998--1999 continuously corrected the inertial navigator using classical gravimetry without GNSS fixes during a four-day interval without resets. Gravity aiding began to control the errors after just over 24 h and, after 48 h, held the latitude and longitude errors below 10\% of the corresponding submarine requirements, although the absolute errors remained classified~\cite{Ryan1999Memphis,Rice2004Geophysical}. Subsequent US Navy work continued to specify the use of gravity-aided navigation for fully submerged submarines~\cite{USNavyGAINS2009}; the navigator specifications and underlying trial data were not disclosed, and the released combined-filter result does not isolate the contribution of gravity-anomaly map matching~\cite{Rice2004Geophysical}.  Similarly, a series of different reports from surface-vessel trials claim root-mean-square errors of 1.92 and 3.19 nautical miles over 34 and 48 h, respectively, a mean error of 915.85 m after approximately 6 h, and errors below 2 nautical miles over 68.8 h~\cite{Wang2016,Wang2024Triangulation,Zhou2026Calibration}.    

It is difficult to ascertain the length to which previous demonstrations of gravity-aided navigation using classical gravimeters relied upon GNSS corrections in their measurement chains.   For instance, submarine trials combined gravity-anomaly matching with gravity-gradient matching and other gravity-derived observations, and open surface-vessel trials have either constructed inertial trajectories from recorded GPS positions or retained GNSS-derived information within the gravimetry toolchain~\cite{Rice2004Geophysical,Wang2016,Zhou2026Calibration}. Zhou \emph{et al.} state that the platform gravimeter used in their trials received GNSS data and used a combined INS/GNSS solution to provide its horizontal datum~\cite{Zhou2026Calibration}, and the Wang studies do not document the kinematic inputs to the gravimeter's E\"otv\"os correction sufficiently to exclude GNSS~\cite{Wang2016,Wang2024Triangulation}. The GNSS-free GravNav demonstration in this work is therefore, as far as we are aware, the only publicly reported end-to-end, GNSS-independent map-matching navigation result from a mobile quantum gravimeter in either gyrostabilization configuration.

The INS hardware used in previous demonstrations is likewise incompletely specified. Zhou \emph{et al.} report gyro-bias and accelerometer-bias accuracies better than $0.003^{\circ}$ h$^{-1}$ and $20\,\mu g$, respectively, which are consistent with representative navigation-grade ranges, but do not identify the INS model~\cite{Zhou2026Calibration,Allen2009IMUGrades}; the other studies do not establish the grade of a measured INS. This traceability is important because mobile gravity recovery is coupled directly to the inertial solution in use. At the trial latitude of approximately $15^{\circ}$ S, an east-velocity error of 1 m s$^{-1}$ produces an E\"otv\"os-correction error of approximately 14 mGal, comparable to or greater than many of the spatial gravity features used for matching; latitude and attitude errors further perturb the correction and the projection of acceleration onto the local vertical. A comprehensive GNSS-free demonstration must therefore document both the exclusion of GNSS from every processing stage and the INS performance that governs these error contributions, as we present in our demonstrations.

Our maritime field trials also include various GNSS-aided mobile gravity surveys, enabling the first direct comparison of gimbaled and strapdown gravimetric survey with the same instrument on the same vessel, and achieving sub-mGal repeatability.  The insights from these surveys are directly relevant to our navigational results as well.  Because the same instrument independently demonstrates recoverable structure at least $50\times$ finer than the map's own resolution, higher-resolution maps generated by this class of gravimeter represent a direct route to tightening the achievable position bound.

In the context of quantum gravimetric survey, classical mobile maritime systems provide a mature performance baseline over spatial scales from several hundred meters to several kilometers~\cite{BellWatts1986,Cai2017Strapdown,Wang2018Strapdown,Yu2020Marine,Yuan2020MarineGravimeters,Johann2021Magnetic,Lu2022Marine}. The best classical strapdown benchmarks of which we are aware remain numerically superior to the results reported here, including 0.64 mGal crossover RMS at 750 m resolution, crossover RMS values of 0.17 and 0.41 mGal for the SAG-2M and SGA-WZ in a side-by-side trial, and approximately 0.26 mGal repeat-line precision after magnetic calibration of an iMAR system~\cite{Yu2020Marine,Yuan2020MarineGravimeters,Johann2021Magnetic}. These values provide scale comparisons rather than a formal ranking because the spatial bandwidths, line geometries, adjustments, and residual normalizations differ among experiments.

It is also instructive to compare the results presented in this work  against previous demonstrations of mobile quantum gravimetric survey.  The 0.67 mGal reported in-run stability, and the 0.16 mGal run-to-run repeatability from the strapdown-only traversals, are numerically close to the 0.9 mGal crossover-derived error reported for the gyro-stabilized (gimbaled) atom gravimeter of Bidel \textit{et al.}, and comparable to its 0.5 mGal forward--backward repeatability and the 0.42--0.46 mGal consistency metrics subsequently reported with other stabilized quantum systems~\cite{Bidel2018,Zhou2024DynamicAtom}. A separately reported strapdown, absolute atomic gravimeter has been validated over multi-week shipboard deployments, including at RIMPAC 2022, with a reported difference from the SS32 satellite gravity-anomaly map of $-0.1\pm2.2$ mGal over a 21-day mission profile~\cite{Ledbetter2025VectorAtomic,AboShaeer2024JQI,VectorAtomic2023}. This is an \emph{external} map-residual metric rather than the \emph{internal} pairwise repeatability values reported here, so the two are not a like-for-like comparison (see \emph{Methods}); nonetheless, our route-wide and strapdown-only pairwise-difference RMS values are both numerically lower than this reported map-residual RMS, and confirm that strapdown quantum gravimetry has independently been fielded at sea. 

%The present result does not, however, improve on the lowest published gimbaled quantum repeatability metric noted above. It is numerically tighter than the only other published strapdown quantum gravimetry figure of comparable kind, an external map-residual RMS of 2.2 mGal (that metric also carries map-truth error absent from our internal pairwise-difference comparison, so the two are not strictly like-for-like). Those disclosures, however, do not separately distinguish in-run stability from run-to-run repeatability, a distinction we report explicitly for both configurations (see \emph{Methods}).  
The core advance we present is that, to our knowledge, this is the first controlled comparison of gimbaled and strapdown gyrostabilization using a single quantum gravimeter: the ONERA benchmarks compared here were obtained exclusively with mechanical stabilization of the sensitive axis~\cite{Bidel2018,Zhou2024DynamicAtom,Gong2026Adaptive}, and other public disclosures of strapdown quantum gravimetry report only that configuration~\cite{Ledbetter2025VectorAtomic,AboShaeer2024JQI,VectorAtomic2023}.  By contrast, Fig.~\ref{fig:gimbal-strapdown} compares both configurations of the same instrument directly over the same traversal with the instrument installed in the same location on the same vessel.

The 450 km road-based survey experiment we report extends our assessment of the robustness of the quantum gravimeter employed here to a faster and more impulsive carrier environment, and provides surprisingly favorable results relative to prior ground-based mobile gravimetry studies in the open literature. The pairwise difference of RMS 6.3 mGal in-run stability is the metric most directly analogous to the definition of "internal consistency" used in several ground-vehicle studies of classical mobile gravimetry~\cite{Yu2015Road,Yu2017Road}. Although the internal consistency of 6.3 mGal is less favorable than the reported 1.2 mGal for SGA-WZ02, our trial is conducted at 100 km h$^{-1}$ over 450 km compared to 40 km h$^{-1}$ over 35 km, and with much more significant elevation change of 750 m compared to 50 m. The closest identified continuously moving terrestrial quantum-atom-gravity experiment operated at 0.055 m s$^{-1}$, more than 500 times slower than the carrier speed used here~\cite{Cheng2022DynamicVehicle}.  We anticipate improved performance at lower speeds with lower RMS heave, in accordance with the results obtained in Fig.~\ref{fig:gimbal-strapdown}(d) and (e), which will open up the application of the quantum sensor used here to practical land-based mobile gravimetry. 

\begin{figure*}[tb]
    \centering
    \includegraphics[width=1.5\columnwidth]{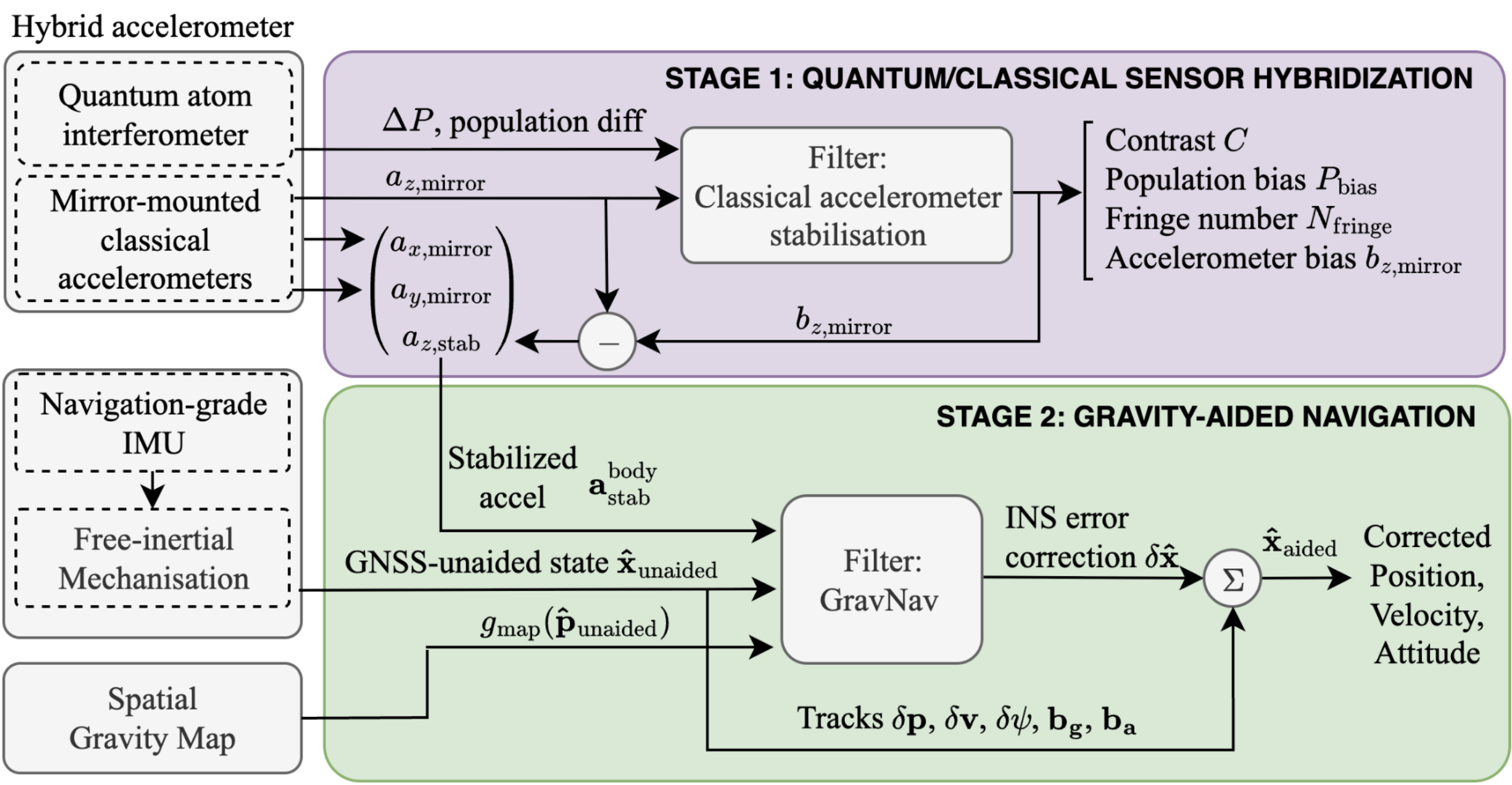}
    \caption{\textbf{Gravity-aided navigation system architecture.}
    In Stage 1, the atom-interferometer population difference $\Delta P$ is combined with the mirror-mounted classical-accelerometer measurements. The stabilization filter estimates the interferometer contrast $C$, population bias $P_{\mathrm{bias}}$, fringe number $N_{\mathrm{fringe}}$ and vertical classical-accelerometer bias $b_{z,\mathrm{mirror}}$. The corrected vertical channel is combined with the transverse classical channels to form the stabilized body-frame acceleration $\mathbf{a}^{\mathrm{body}}_{\mathrm{stab}}$. In Stage 2, the GravNav filter combines this acceleration with the GNSS-unaided inertial state $\hat{\mathbf{x}}_{\mathrm{unaided}}$ and the gravity-map value $g_{\mathrm{map}}(\hat{\mathbf{p}}_{\mathrm{unaided}})$. The filter tracks position, velocity and attitude errors and inertial-sensor biases, estimates the correction $\delta\hat{\mathbf{x}}$, and produces the aided state $\hat{\mathbf{x}}_{\mathrm{aided}}$.}
    \label{fig:gravnav-architecture}
\end{figure*}

The first priority for future work is to extend these demonstrations to a long-duration, fully closed navigation campaign. Extending the present 6 h GNSS-free navigation experiment through multi-week operation is necessary to test whether atom-referenced bias stability produces an endurance advantage after the gravity ties, empirical drift models and crossover corrections available to classical systems are taken into account. Stationary measurements provide a positive indication that long endurance missions are possible, as they demonstrate device stability over an uninterrupted 56 h record (see \emph{Methods}). 

A second priority is reducing total-system size, weight and power (SWaP). The present trials provide two encouraging indications: strapdown operation removes the active gimbal, and no dedicated external thermal enclosure was required. A meaningful operational comparison must nevertheless include the complete sensor head, vacuum hardware, lasers, control electronics and power supply. The 33--134 L packaged volumes reported for high-performing classical maritime strapdown systems therefore remain the relevant system-level benchmark~\cite{Johann2023Thesis,Wang2018SGAWZ02,Wang2018Strapdown}. Reducing a complete quantum atom gravimeter below this range is a concrete near-term target.  Q-CTRL has already produced next-generation mobile quantum-gravity systems leveraging new forms of software-ruggedization in lieu of mechanical stabilization countermeasures and has demonstrated in-motion lab-based performance superior to that reported here in much smaller SWaP envelopes. Realizing the full potential of such systems will open deployment across a broader class of unmanned platforms, extending quantum sensing to navigation, gravity surveying, and anomaly detection in all domains.

\section*{Methods}

\subsection*{Maritime and land traversal details}

In March 2026, Q-CTRL chartered the 29 m research vessel \textit{MV Miss Rankin} (Fig.~\ref{fig:marine-trial}d) for a week-long campaign in the Coral Sea north of Cairns, Australia. The 280 km coastal route shown in Fig.~\ref{fig:gravity-surveys}c was traversed northbound and southbound with the sensor actively gimbaled and then repeated in both directions with the gimbal disabled, providing the four registered profiles analyzed over a common 280 km line in Fig.~\ref{fig:gimbal-strapdown}. The gimbaled traversals followed a tropical low and encountered Sea States 2--3; the vessel attitude repeatedly exceeded the gimbal's angular travel. The strapdown traversals were conducted in calmer conditions, typically Sea State 1. The vessel autopilot held deviations from the programmed route to typically less than 10 m, and the typical underway speed was 8 knots.

Over the Sea State 2 conditions sampled during the repeated coastal traversals, the vessel exhibited lateral and vertical accelerations and rotation rates of 7 mg RMS ($0.10\,g$ peak-to-peak), 6 mg RMS ($0.12\,g$ peak-to-peak) and $0.37^{\circ}$ s$^{-1}$ RMS ($7.1^{\circ}$ s$^{-1}$ peak-to-peak), respectively, in strapdown operation, and 12 mg RMS ($0.13\,g$ peak-to-peak), 20 mg RMS ($0.29\,g$ peak-to-peak) and $0.22^{\circ}$ s$^{-1}$ RMS ($4.4^{\circ}$ s$^{-1}$ peak-to-peak) in gimbaled operation.

The 145 km Great Barrier Reef loop in Fig.~\ref{fig:gravity-surveys}c was traversed three times, with conditions outside the reef typically reaching Sea State 4. The sensor was also operated overnight on its gimbal while the vessel was anchored in Trinity Inlet, Cairns, where conditions were typically Sea State 1. Together, the coastal, loop and anchored measurements sampled gimbaled and strapdown operation across calm, moderate and rough vessel-motion conditions. Across the full trial, including the Sea State 4 conditions sampled during the loop-route traversals, the vessel experienced roll angles of up to $\pm12^{\circ}$ and roll rates of up to $10^{\circ}$ s$^{-1}$.

For the land trial, the same sensor head was operated on its gimbal inside a van without specialized modification of the vehicle platform. Northbound and southbound profiles were acquired along a 450 km section of the Hume Highway between south-west Sydney and Mullengandra, New South Wales (Fig.~\ref{fig:road-gravity}), at an average speed of 100 km h$^{-1}$ (27.8 m s$^{-1}$). Kilometer zero was defined at the north-eastern end of the route.

\subsection*{Quantum sensor head}

The quantum sensor head installed in the vessel cabin is shown in Fig.~\ref{fig:marine-trial}c. It contains a source of cold $^{87}$Rb atoms, three mutually orthogonal atom-interferometer axes and supplementary classical inertial sensors in an approximately 30 L package; the laser and control electronics are housed in the adjacent rack visible in the same panel. Only one atom-interferometer axis, corresponding to the nominally vertical body-axis channel, is used in the field trials reported here. Atoms are collected and cooled in a magneto-optical trap at the center of the sensor-head vacuum chamber. Polarization-gradient cooling and state preparation transferred the cloud to the $\lvert F=1,m_F=0\rangle$ ground state. 

An inertially sensitive atom interferometer sequence is then produced by dynamically switching the frequency and polarization of the magneto-optical-trap beams. The resulting Mach--Zehnder interferometer uses three square $\pi/2$--$\pi$--$\pi/2$ pulses to split, redirect and recombine the freely falling matter waves during a total time $2T$, with $T$ typically ranging from 3 ms to 20 ms (Fig.~\ref{fig:marine-trial}a). The recombined interfering matter waves are measured by fluorescence detection collected by photodiodes through two lenses mounted inside the vacuum chamber. The magneto-optical-trap beams and magnetic fields are then restored immediately, recapturing and recycling the atomic cloud for the next measurement. This sequence supports cycle rates up to 20 Hz, primarily limited by the atomic free-fall time. The fiber-coupled laser beam is retro-reflected through the atoms from a mirror carrying a rigidly mounted three-axis classical accelerometer (Fig.~\ref{fig:marine-trial}b), sampled at up to 8 kSa s$^{-1}$ and providing the common mechanical reference used by the hybridization stage.

\subsection*{Quantum--classical sensor hybridization}

The first processing stage is common to the navigation and gravity-survey branches (Figs.~\ref{fig:gravnav-architecture} and \ref{fig:gravity-signal-processing}). The atom interferometer supplies the normalized output-population difference $\Delta P$, while the mirror-mounted classical accelerometer supplies the three body-frame components $a_{x,\mathrm{mirror}}$, $a_{y,\mathrm{mirror}}$ and $a_{z,\mathrm{mirror}}$. The classical channel provides the bandwidth and dynamic range required to follow platform motion and to resolve the atom interferometer's $2\pi$ phase ambiguity; the atom measurement supplies the low-frequency reference used to stabilize the classical bias. This division of bandwidth follows established hybrid quantum--classical accelerometer methods~\cite{Lautier2014Hybrid,Cheiney2018Hybrid,Bidel2018}.

The full-rate classical-accelerometer data, sampled at 8 kSa s$^{-1}$, is used for vibration compensation of the atom interferometer. For downstream navigation-centric processing, this stream is decimated to 300 Sa s$^{-1}$, matching the native sample rate of the INS described below.

A stabilization filter compares $\Delta P$ with the atom-response-function-weighted vertical classical acceleration, following the direct phase-reconstruction strategy used in dynamic atom gravimetry~\cite{Bidel2018,Bidel2020Airborne,Bidel2023Airborne}. At each atom-interferometer cycle, the filter jointly tracks the interferometer contrast $C$, population offset $P_{\mathrm{bias}}$, integer fringe number $N_{\mathrm{fringe}}$ and vertical classical-accelerometer bias $b_{z,\mathrm{mirror}}$. Subtracting the estimated bias from $a_{z,\mathrm{mirror}}$ produces the stabilized vertical acceleration $a_{z,\mathrm{stab}}$. This channel is combined with the two transverse classical measurements to form the stabilized body-frame vector $\mathbf{a}^{\mathrm{body}}_{\mathrm{stab}}=(a_{x,\mathrm{mirror}},a_{y,\mathrm{mirror}},a_{z,\mathrm{stab}})^{\mathsf T}$, which is passed to either of the two second-stage processing branches below.

For comparison, a classical-only gravity estimate is formed using the identical processing chain but omitting the atom-interferometer bias correction. This branch retains the raw vertical mirror-accelerometer channel $a_{z,\mathrm{mirror}}$ in place of $a_{z,\mathrm{stab}}$, isolating the contribution of the atom-referenced bias correction to the resulting gravity estimate.

\begin{figure}[tb]
    \centering
    \includegraphics[width=\columnwidth]{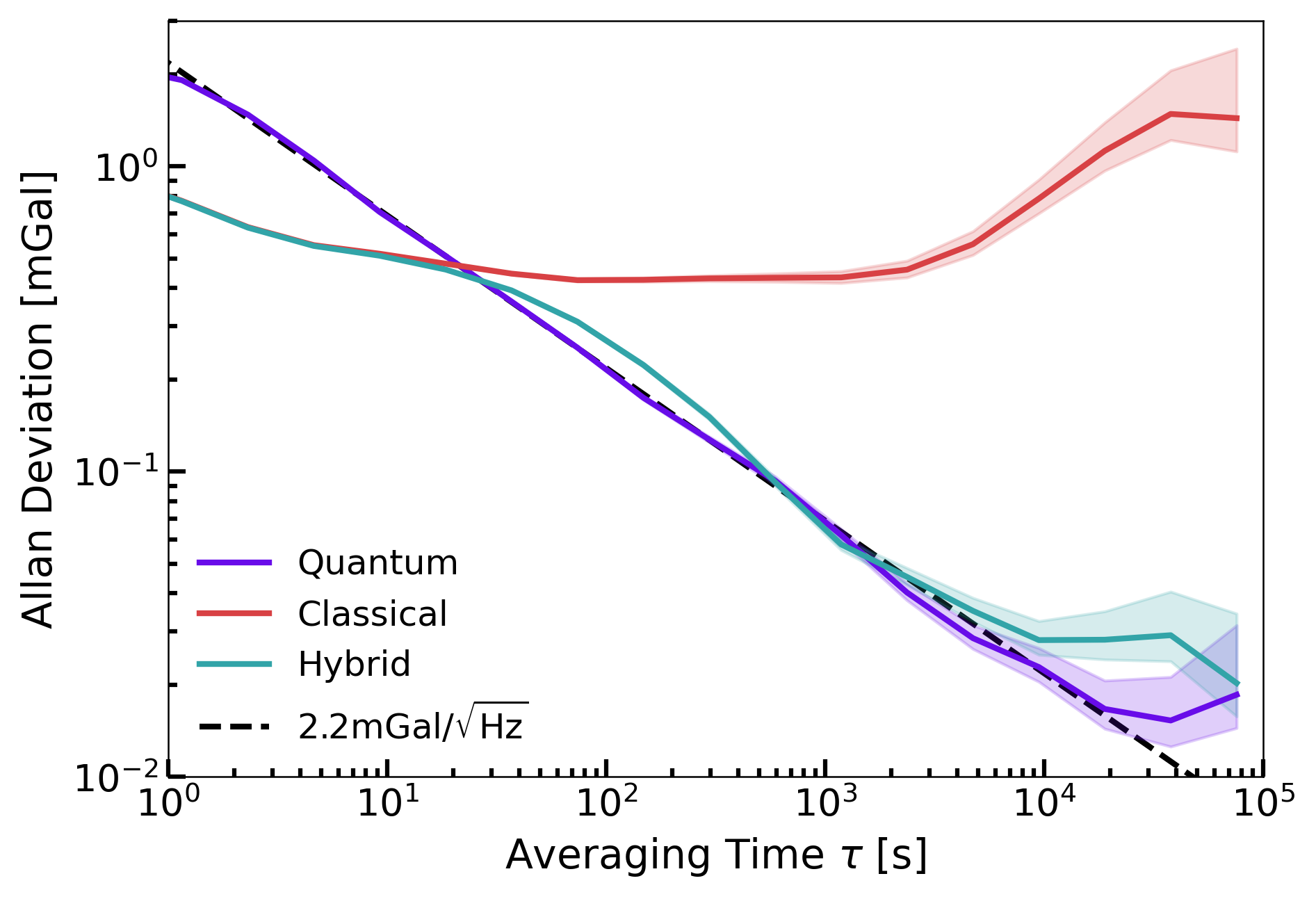}
    \caption{Long-term stability of the quantum gravimeter under stationary laboratory conditions.
    Allan deviation of the quantum (purple), classical (red) and hybrid (teal) gravity estimates calculated from an uninterrupted 56 h stationary record and plotted versus averaging time $\tau$. The dashed black line shows $\tau^{-1/2}$ scaling with a fitted one-second coefficient of 2.2 mGal Hz$^{-1/2}$. Shaded bands denote the spread of the Allan-deviation estimates. }
    \label{fig:stability}
\end{figure}

We demonstrate the quantum--classical sensor hybridization operating as intended in Fig.~\ref{fig:stability}, where the Allan deviation tracks a single uninterrupted 56 h stationary record in laboratory conditions. At $\tau=1$ s, the classical and hybrid estimates overlap at approximately 0.78 mGal, below the quantum estimate of 1.94 mGal, showing that the hybrid retains the lower short-term noise of the classical accelerometer. The quantum estimate subsequently follows $\tau^{-1/2}$ averaging with a fitted coefficient of 2.2 mGal Hz$^{-1/2}$ and crosses below the classical estimate after approximately 20 s. The classical Allan deviation reaches a floor near 0.42 mGal before rising to 1.5 mGal at $4\times10^{4}$ s, whereas the atom-referenced hybrid remains between approximately 0.020 and 0.028 mGal from $10^{4}$ to $8\times10^{4}$ s. At $8\times10^{4}$ s, the hybrid value of approximately 0.020 mGal is therefore about 70 times lower than the classical value and within approximately 10\% of the quantum value, showing that the hybrid transitions from classical short-term sensitivity to atom-referenced long-term stability. This behavior is consistent with hybrid atom gravimeters previously deployed on mobile marine platforms, which reported stationary sensitivities of 0.8 and 0.447 mGal Hz$^{-1/2}$~\cite{Bidel2018,Che2022ShipborneHybrid}. The uninterrupted 56 h record, over which the quantum estimate continues to average down before remaining near 0.02 mGal at the longest averaging times, demonstrates the low intrinsic drift of the atom reference over the measurement interval.

\subsection*{Gravity-aided navigation}

The complete gravity-aided navigation demonstration (Results, ``Quantum navigation with gravity map matching'') uses the 45 nmi (83 km) southbound-coastal-route segment (``Field trials'').

Mobile gravity recovery is coupled to the navigation state because the measured specific force must be reduced to the local vertical and corrected for the apparent acceleration produced by motion over the rotating Earth. The dominant first-order E\"otv\"os term has sensitivity $\partial a_{\mathrm{E}}/\partial v_{\mathrm{East}}\simeq 2\Omega_{\oplus}\cos\varphi_{\mathrm{lat}}$. This contributes an illustrative error more than an order of magnitude larger than the route-wide repeatability reported in Fig.~\ref{fig:gimbal-strapdown}b and is comparable to many of the spatial gravity features used for matching. Latitude error perturbs the E\"otv\"os coefficient, while attitude error projects horizontal vessel acceleration into the estimated down-axis acceleration. In the absence of GNSS these quantities are supplied by the drifting inertial solution, motivating their joint estimation with the gravity-derived navigation correction rather than treating gravity recovery and inertial aiding as independent operations.

% At the approximately $15^{\circ}$ S operating latitude, an east-velocity error of 1 m s$^{-1}$ would therefore produce an E\"otv\"os-correction error of approximately $1.4\times10^{-4}$ m s$^{-2}$, or 14 mGal. 

The inertial instrument used in both processing modes is an Advanced Navigation Boreas D90 FOG GNSS/INS. The unit contains three digital fiber-optic gyroscopes and three navigation-grade quartz accelerometers. The manufacturer specifies accelerometer bias instability, noise density and velocity random walk of $7\,\mu g$, $30\,\mu g\,\mathrm{Hz}^{-1/2}$ and $17$ mm s$^{-1}$ h$^{-1/2}$, respectively, and gyroscope bias instability, noise density and angle random walk of $0.001^{\circ}$ h$^{-1}$, $0.06^{\circ}$ h$^{-1}$ Hz$^{-1/2}$ and $0.001^{\circ}$ h$^{-1/2}$, respectively~\cite{AdvancedNavigationBoreasD90}. For the gravity-aided navigation demonstration, Q-CTRL independently mechanizes the recorded raw three-axis angular-rate and specific-force measurements. The position, velocity and attitude solution produced by the Boreas embedded INS, including its GNSS aiding, is not used. Without GNSS or gravity aiding, the independently mechanized free-inertial position solution drifts at an average rate of approximately 5 km h$^{-1}$ over the 5 h evaluation window, motivating the gravity-based aiding described below.

The resulting free-inertial state $\hat{\mathbf{x}}_{\mathrm{unaided}}$ is combined with $\mathbf{a}^{\mathrm{body}}_{\mathrm{stab}}$ in the GravNav stage (Fig.~\ref{fig:gravnav-architecture}). The spatial gravity map was evaluated using the current unaided position estimate, giving $g_{\mathrm{map}}(\hat{\mathbf{p}}_{\mathrm{unaided}})$. The GravNav error-state filter compares the recovered gravity information with the mapped field and tracked position, velocity and attitude errors $(\delta\mathbf{p},\delta\mathbf{v},\delta\boldsymbol{\psi})$, together with gyroscope and accelerometer biases $(\mathbf{b}_{g},\mathbf{b}_{a})$. Its estimated state correction $\delta\hat{\mathbf{x}}$ is applied to the unaided solution to generate the gravity-aided position, velocity and attitude estimate $\hat{\mathbf{x}}_{\mathrm{aided}}$. 

As mentioned above, GNSS observations are excluded from this complete processing branch. No GNSS-derived position, velocity or attitude entered the independent inertial mechanization, hybrid acceleration, gravity recovery, map evaluation, GravNav filter or state correction. GNSS data are logged independently and used only after the experiment to construct the withheld reference trajectory and evaluate the errors shown in Fig.~\ref{fig:gravnav}.

\subsection*{Gravity signal for surveying and anomaly detection}

\begin{figure*}[thb]
    \centering
    \includegraphics[width=1.5\columnwidth]{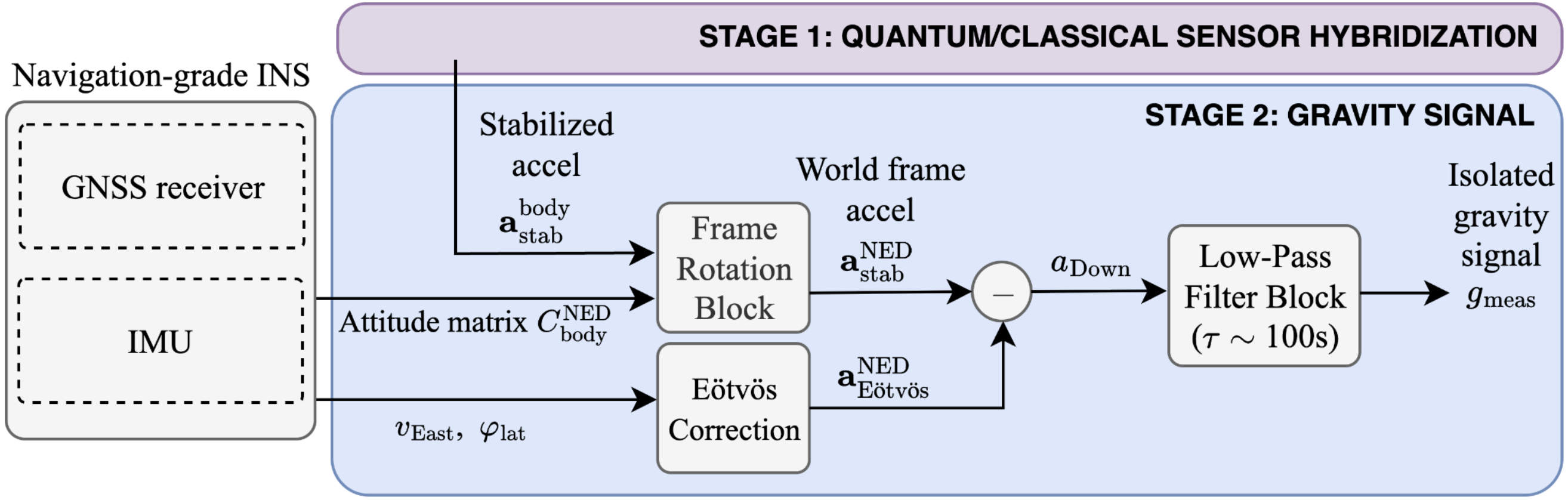}
    \caption{\textbf{Hybrid-accelerometer processing and gravity-signal extraction.}
    Stage 1 supplies the stabilized body-frame acceleration $\mathbf{a}^{\mathrm{body}}_{\mathrm{stab}}$. In Stage 2, the attitude matrix $\mathbf{C}^{\mathrm{NED}}_{\mathrm{body}}$ from the GNSS-disciplined Boreas D90 solution rotates this acceleration into the North--East--Down frame. The E\"otv\"os correction is calculated using eastward velocity $v_{\mathrm{East}}$ and latitude $\varphi_{\mathrm{lat}}$ from the same solution and removed from the down-axis acceleration. A fourth-order Bessel low-pass filter with a characteristic time constant of order 100 s isolates the measured gravity signal $g_{\mathrm{meas}}$.}
    \label{fig:gravity-signal-processing}
\end{figure*}

The independent gravity-survey branch uses the same stabilized body-frame acceleration but, unlike the GravNav branch, uses the GNSS-disciplined position, velocity and attitude solution from the Boreas D90 (Fig.~\ref{fig:gravity-signal-processing}). The attitude matrix $\mathbf{C}^{\mathrm{NED}}_{\mathrm{body}}$ rotated $\mathbf{a}^{\mathrm{body}}_{\mathrm{stab}}$ into the North--East--Down frame; in strapdown operation, this attitude-based rotation is the software analogue of the mechanical gyrostabilization otherwise provided by the gimbal. Eastward velocity and latitude from this solution are used to calculate the E\"otv\"os correction, which is removed from the down-axis acceleration; its position is also used to georeference the resulting gravity profile. Vessel heave and other platform accelerations are suppressed with a fourth-order Bessel low-pass filter. This filter architecture is established in mobile maritime atom gravimetry, where its time constant is selected as a trade-off between rejection of vessel vertical acceleration and along-track spatial resolution~\cite{Bidel2018,Zhou2024DynamicAtom}. A representative time constant was approximately 100 s; the maritime analyses use values from 60 to 1200 s, as reported in the corresponding figure captions. The output of this stage is the isolated gravity signal $g_{\mathrm{meas}}$ used for the survey-map comparisons and spatial-anomaly analysis in Figs.~\ref{fig:gravity-surveys}, \ref{fig:gimbal-strapdown} and \ref{fig:road-gravity}. This GNSS-dependent surveying branch is processed separately from, and supplies no observations to, the GNSS-free GravNav result, following the routine survey practice described above (Results, ``Repeated routes recover fine gravity anomalies'').

In Fig. \ref{fig:gimbal-strapdown}(d), the stability of the sensor when subject to out-of-band platform motion is estimated by looking at the RMS of the ensemble-averaged pairwise differences of gravity traces, $\sigma_k=\sqrt{\frac{1}{2} \langle \mathrm{Var}_k(g_i-g_j) \rangle}$, over different time windows, which are indexed by $k$. This RMS value provides an estimate of the in-run stability of the sensor as a function of the platform heave. Note that for computing the weighted average, the northbound gimbaled data uses an equal weighting of each of the other three traces (southbound gimbal and the strapdown traces). The other three traces, however, do not include the pairwise difference against the northbound gimbaled data owing to the level of noise.

In Fig. \ref{fig:gimbal-strapdown}(e), the noise floor of the sensor stack is estimated by examining the noise spectral density of the residual between each trace and a ground-truth estimation constructed from the ensemble average of the remaining gravity traces. The residual for the northbound gimbal gravity trace is found by taking the difference between this trace and the average of the southbound gimbal and strapdown traces. The noise floor is then estimated by computing the RMS of the noise spectral density below 0.01 Hz. Note that similar to Fig. \ref{fig:gimbal-strapdown}(d), the ground truth traces estimated for the southbound gimbal data and the strapdown traces omit the northbound gimbal data owing to its noise. These estimates are made using gravity traces with a 60 s low-pass filter time-constant, to better preserve the close-to-DC noise.

\section*{Acknowledgments}

Q-CTRL acknowledges support for this trial from the UK Royal Navy via the Defence and Security Accelerator (DASA), and the contributions of many team members developing hardware and software supporting these efforts.

\bibliography{references.bib}

\end{document}